\documentclass[preprint,prX,amsmath, amssymb]{revtex4}
\usepackage{graphicx}% Include figure files
\usepackage{dcolumn}% Align table columns on decimal point
\usepackage{bm}% bold math
\usepackage{setspace}

\begin{document}

\preprint{APS/123-QED}

\title{Efficient Evaluation of Casimir Force in Arbitrary Three-dimensional Geometries by Integral Equation Methods}% Force line breaks with \\

\author{Jie L. Xiong}
 %\altaffiliation[Also at ]{Faculty of Engineering, University of Hong Kong.}%Lines break automatically or can be forced with \\
\author{Mei Song Tong}
\author{Phillip Atkins}
\author{Weng Cho Chew}%

%\begin{gather}
	
%\end{gather}\footnote{wcchew@hku.hk}
 \email{wcchew@hku.hk}
\affiliation{%
Faculty of Engineering, University of Hong Kong, Hong Kong;
Electrical and Computer Engineering, University of Illinois, Urbana-Champaign, Illinois, USA, 61801}%

\renewcommand{\vec}[1]{\mbox{${\bf #1}$}}
\renewcommand{\v}[1]{{{\bf #1}}}
\newcommand{\dyad}[1]{\mbox{$\overline{ \v{#1} }$}}
\newcommand{\pd}[2]{\frac{\partial #1}{\partial #2}}
\def\prs{\v{r}, \v{r}'}
\def\prhos{\bs \rho, \bs \rho'}
\def\bs{\boldsymbol}
\def\tG{\widetilde{G}}
\def\tE{\widetilde{E}}
\def\tH{\widetilde{H}}
\def\grr{g(\v{r},\v{r'})}
\def\bs{\boldsymbol}
\date{\today}% It is always \today, today,
             %  but any date may be explicitly specified

\begin{abstract}
In this paper, we generalized the surface integral equation method for the evaluation of Casimir force in arbitrary three-dimensional geometries. Similar to the two-dimensional case, the evaluation of the mean Maxwell stress tensor is cast into solving a series of three-dimensional scattering problems. The formulation and solution of the three-dimensional scattering problem is well-studied in classical computational electromagnetics. This paper demonstrates that this quantum electrodynamic phenomena can be studied using the knowledge and techniques of classical electrodynamics.
\end{abstract}

%\pacs{Valid PACS appear here}% PACS, the Physics and Astronomy
                             % Classification Scheme.
%\keywords{Suggested keywords}%Use showkeys class option if keyword
                              %display desired
\maketitle

%\section{Introduction}

%As recent advances in miniaturization process leads to the development of micro- and nano-electromechanical systems (MEMS and NEMS), it is also increasingly of practical interest to study the role and influence of Casimir force in these small devices now. 

The Casimir force has first been predicted by Casimir in 1948 \cite{Casimir_1948}. It exists among charge-neutral bodies due to the quantum fluctuation of the elecromagnetic fields in vacuum. Although it is tiny when the objects have large separations, it becomes the dominant force between charge-neutral objects when the separation is below micron meters. 

Micro-electromechanical systems (MEMS) are micron-sized devices
in which mechanical elements and moving parts, such as tiny sensors and actuators are carved into a silicon substrate. As further 
miniaturization takes place, the devices may reduce to nano-scale and become nanoelectromechanical systems (NEMS). They have a wide class of applications. For example, the release of the airbag in cars is controlled by a MEMS-based accelerometer.
One of the principal causes of malfunctioning in MEMS is stiction, i.e., the collapse of movable elements into nearby surfaces, resulting in their permanent adhesion. Casimir effect is often an important underlying mechanism causing this phenomenon \cite{Buks_2001}. On the other hand, this effect could be put into good use. Capasso and his group showed that the force can be used to control the actuation of a micromachined torsional device \cite{Chan_2001sci}. This group also showed that the Casimir attraction can be used to make a nonlinear oscillator. The force influences the dynamical properties of a micromachined device, changing its resonance frequency, hysteretic behavior, and bistability in its frequency response to an AC excitation. They proposed that this device could serve as a position sensor working in nanometer scale \cite{Chan_2001prl, Capasso_munday_2007}.

Since Casimir effect is rather significant in MEMS and NEMS devices, it is important to be able to predict it accurately before the devices going into manufacturing. The geometry of the MEMS and NEMS devices may be complicated; thus approximation methods, such
as the proximity force approximation (PFA) \cite{Bordag_2001},
and optical approach in terms of virtual photons moving along ray
optical paths \cite{Scardicchio_2005}, are no longer sufficient. Instead, an exact and general numerical method that can handle arbitrary geometry is desired. With the numerical method, we can get the desired accuracy by giving enough computational resource. 
Recently, major progress has been made in the development of numerical methods for Casimir force. In general, there are two popular approaches: the path integral approach (also known as EGJK method) and the Maxwell stress-tensor approach. 

Schwinger is the first one to attribute the Casimir interaction to fluctuating current and charge densities inside the objects. 
%The path integral method was proposed by B\"uscher \cite{Buscher_2005}. It begins by expressing the Casimir energy as the logarithm of a functional integral over all field fluctuations constrained by the boundary conditions on a set of surfaces. Then the functional integral is performed over the field. 
With path integral method, the Casimir energy is expressed as the logarithm of a functional integral over all allowed configurations of the fluctuating currents, weighted by the appropriate action \cite{Buscher_2005}. It has been generalized in \cite{EGJK_2007} (EGJK) for compact objects of arbitrary shape and separation and applied to predict the force between a cylinder and a plate \cite{Emig_2006}, and between two spheres \cite{EGJK_2007}. EGJK's agorithm is efficient to predict the Casimir energy between compact 3D objects of spheroidal or nearly-spheroidal shape. The algorithm leads to analytically tractable series solution. But it is not of practical use for general geometries due to the poor convergence rate for objects with corners. This limitation occurs since they used spherical basis functions to expand the current distribution. It could be overcome by expanding the currents with basis functions defined on a pair of adjacent triangles, which is known as the RWG basis \cite{Reid_2009, RWG}. The RWG basis is vastly popular in the surface integral method in computational electromagnetics. After some manipulation, the evaluation of Casimir energy depends only on solving for all the eigenvalues of the impedance matrix obtained from the integral equation by the method of moments. 

The theoretical foundation for the Maxwell stress tensor approach was built by Lifshitz and his coworkers \cite{dzyaloshinskii_1961}: the net Casimir force on a body can be expressed as an integral over any closed surface around the body of the mean electromagnetic stress tensor, integrated over all frequencies. And the fluctuation-dissipation theorem states that the mean fluctuating field, which forms the Maxwell stress tensor, is related to the imaginary part of the dyadic Green's
function of the same geometry. The dyadic Green's function can be
evaluated by existing numerical methods in computational
electromagnetics (electrodynamics) \cite{Rodriguez_2007pra}. Rodriguez has demonstrated the applicability of this method by using a simple finite-difference frequency-domain (FDFD) method to calculate the dyadic Green's function in $z$-invariant structures \cite{Rodriguez_2007prl, Rodriguez_2008}. This approach relates the quantum electrodynamic phenomena with classical electromagnetics which motivates our piece of work: to evaluate the Casimir force among arbitrary objects using integral equation method. With surface integral equation method, the number of unknowns can be greatly reduced. Moreover, fast algorithms \cite{book_chew_2001} are available to further improve the efficiency of the method. The computational complexity of both the stress tensor approach and the path integral approach can be reduced to $O(N\log N)$ times the number of iterations at best, $N$ being the number of surface unknowns. Thus their efficiency is comparable. The main difference between them is the physical quantity being calculated directly: path integral method calculates the energy while stress tensor approach calculates the force distribution (pressure).

%In this paper we propose to use the integral equation method to calculate the dyadic Green's function of a general geometry and its derivatives in order to get the Casimir force. The dyadic Green's function is obtained by solving the electromagnetic problem with a monopole point source placed very close to the surface of the object, and then evaluating the scattered field at the same location. Compared to finite difference methods, integral equation methods involve a discretization only at the surface of the coupled objects, thus lead to fewer unknowns. 

In a paper published earlier \cite{Jie_2009}, we have demonstrated that surface integral equation method could be applied to evaluate the dyadic Green's function and its derivatives in two-dimensional (2D) geometries. By using this method, the Casimir force between 2D objects could be simulated using much fewer unknowns and can be easily accelerated by fast-algorithms readily developed in computational electromagnetics. In this paper,  we would extend the surface integral equation method for arbitrary three-dimensional (3D) geometries. The idea is similar to the last paper \cite{Jie_2009}: the Casimir force is expressed in terms of the dyadic Green's function and its derivatives and they could be calculated from the integral equation by using different types of dipole point sources.

%\section{Formulation}	
%First, we would review the fundamental theory and key steps of calculating Casimir force from the Maxwell stress tensor and dyadic Green's function, which has been given in \cite{Rodriguez_2007pra}. 
The starting point of the stress tensor approach is that the net Casimir force acting on the surface $S$ of an object is given by an surface integral of the mean stress tensor on it \cite{Rodriguez_2007pra}:
\begin{eqnarray}\label{eqn:c1}
\vec{F} = \oint_S \langle \dyad{T} (\v r') \rangle \cdot d\v s'
\end{eqnarray}
where $\dyad{T}$ is the Maxwell stress tensor defined as in \cite{jackson_1975}.
For an arbitrary 3D object, we define $\hat{u}$ and $\hat{v}$ as the tangent vector at the surface, and $\hat{n}$ as the normal direction of the surface. They satisfy the relationship $\hat{u} \times \hat{v} = \hat {n}$. 
%The stress tensor and dyadic Green's function at point $\v r_i$ is conveniently expressed in terms of $\hat u$, $\hat v$ and $\hat n$ at this point as 
%\begin{eqnarray} \label{eqn:d6}
%\v F(\v r_i) = \langle T_{nn} (\v r_i) \rangle \hat n_i  + \langle T_{un}(\v r_i) \rangle \hat u_i + \langle T_{vn}(\v r_i) \rangle \hat v_i
%\end{eqnarray}
If the object is a perfect conductor, the tangential electric field $E_u$, $E_v$ and the normal magnetic field $B_n$ vanishes at the surface of the object. The stress tensor could be simplified as follows: 
\begin{eqnarray} 
T_{nn}(\v r) &=& \frac{1}{2}\epsilon_0 E^2_n(\v r) - \frac{1}{2\mu_0} B^2_u (\v r) - \frac{1}{2\mu_0} B^2_v (\v r) \label{eqn:d7}\\
T_{un}(\v r) &=& T_{vn}(\v r) = 0 \hspace{0cm}
\end{eqnarray}
Thus for perfect conductors, the Casimir force becomes a pressure on the surface.

%\begin{eqnarray}\label{eqn:c2}
%T_{ij} &=& \epsilon_0 E_i E_j + \frac{1}{\mu_0}B_i B_j  \nonumber \\ && \hspace{1cm} - \frac{1}{2}\left(\epsilon_0 \sum_{k=1}^3 E^2_k + \frac{1}{\mu_0} \sum_{k=1}^3 B^2_k\right)\delta_{ij}
%\end{eqnarray}
The average of the fluctuating electric and magnetic fields in the ground state is obtained from the fluctuation-dissipation theorem \cite{landau_1980}:
\begin{eqnarray}
\langle 0| \hat{E}_i(\v{r}, t)\hat{E}_j (\v{r}', t) |0 \rangle 
&=& \frac{\hbar}{\pi} \text{Im}\int_0^{\infty}   \omega^2 {G}_{ij}(\v{r}, \v{r}', \omega) d\omega \label{eqn:c3} \\
\langle 0| \hat{B}_i(\v{r}, t)\hat{B}_j (\v{r}', t) |0 \rangle &=& \frac{\hbar}{\pi} \text{Im}\int_0^{\infty}  (\nabla \times )_{il} (\nabla' \times)_{jm} G_{lm}(\prs, \omega)  d\omega  \label{eqn:c4}
\end{eqnarray}
where the dyadic Green's function is the same as that defined in classical electromagnetics to relate the current to the field in an arbitrary geometry \cite{Jie_2009}.
%where the dyadic Green's function $\dyad{G}(\prs, \omega) $ satisfies the following equation
%%\begin{eqnarray} 
%%\nabla \times \nabla \times {\dyad{G}}(\prs, \omega) - k^2 {\dyad{G}}(\prs \omega) &=& \mu_0 \dyad{I} \delta(\v{r},\v{r}') \label{eqn:c5}
%%\end{eqnarray}
%%The integrand in \eqref{eqn:c3} and \eqref{eqn:c4} are highly oscillatory. {\it Wick rotation} \cite{Rodriguez_2007pra} could be used to deform the integration path from the real axis to the imaginary axis and accelerate the evaluation of the integral significantly.

In 3D structures, the field components and the dyadic Green's function satisfy the following relationship:
\begin{eqnarray}
E_\alpha( \v r) &=&  i\omega   \int_S d\v r' {G}_{\alpha \beta}(k, \v r, \v r') {J}_\beta( \v r') \label{eqn:c31} \hspace{0cm}\\
%\end{eqnarray}
%\begin{eqnarray}
H_u(  \v r) &=& \frac{1}{\mu_0}   \int_S d\v r' \left[\pd{}{v}G_{n\beta}(k,  \prs) - \pd{}{n} G_{v\beta}(k,\prs) \right] {J}_\beta(\v r')  \label{eqn:c32}
\end{eqnarray}
Substituting \eqref{eqn:c31} and \eqref{eqn:c32} into \eqref{eqn:c3} and \eqref{eqn:c4}, 
the average of each term in the stress tensor could be represented by
%\begin{eqnarray}
%\frac{1}{2}\epsilon_0 \langle E_n(\v r) E_n(\v r') \rangle &=& \frac{\hbar \epsilon_0}{2\pi} \text{Im} \int_0^\infty \omega^2 G_{nn}(\v r, \v r', \omega) d\omega =   \frac{\hbar c_0}{2\pi \mu_0} \text{Im} \int_0^\infty \frac{k^2_0}{i\omega} E_{n}(k_0, \v r, s_1) dk_0 \label{eqn:c33} \nonumber \\
%\frac{1}{2\mu_0}\langle B_u(\v r) B_u(\v r') \rangle &=& \frac{\hbar c_0}{2\pi \mu_0} \text{Im} \int_0^\infty dk_0 \left[\pd{^2}{n\partial n'} G_{vv}(\v r,\v r') - \pd{^2}{v\partial n'} G_{nv}(\v r,\v r') \right. \nonumber \\ &&\hspace{1cm}\left. - \pd{^2}{n\partial v'} G_{vn}(\v r,\v r') + \pd{^2}{v\partial v'} G_{nn}(\v r,\v r') \right] \nonumber \\
%&=& \frac{\hbar c_0}{2\pi} \text{Im} \int_0^\infty d(k_0) \left[-H_u(k_0, \v r, s_2) + H_u(k_0, \v r, s_3)\right] \label{eqn:c34}\\
%\frac{1}{2\mu_0}\langle B_v(\v r) B_v(\v r') \rangle &=& \frac{\hbar c_0}{2\pi} \text{Im} \int_0^\infty d(k_0) \left[-H_v(k_0, \v r, s_4) + H_u(k_0, \v r, s_5)\right] \label{eqn:c35}
%\end{eqnarray}
\begin{eqnarray}
\frac{1}{2}\epsilon_0 \langle E_n(\v r) E_n(\v r') \rangle &=& \frac{\hbar c_0}{2\pi \mu_0} \text{Im} \int_0^\infty \frac{k^2_0}{i\omega} E_{n}(k_0, \v r, s_1) dk_0 \label{eqn:c33}  \\
\frac{1}{2\mu_0}\langle B_u(\v r) B_u(\v r') \rangle &=& \frac{\hbar c_0}{2\pi} \text{Im} \int_0^\infty dk_0 \left[-H_u(k_0, \v r, s_2) + H_u(k_0, \v r, s_3)\right] \label{eqn:c34}\\
\frac{1}{2\mu_0}\langle B_v(\v r) B_v(\v r') \rangle &=& \frac{\hbar c_0}{2\pi} \text{Im} \int_0^\infty dk_0 \left[-H_v(k_0, \v r, s_4) + H_u(k_0, \v r, s_5)\right] \label{eqn:c35}
\end{eqnarray}
Here, $\v E(k_0, \v r, s_i)$ and $\v H(k_0, \v r, s_i)$ are the electric/magnetic field at location $\v r$ in the geometry under the excitation of source $s_i$. Their corresponding relationships are summarized in Table \ref{table:d1}. 

\begin{table}[htbp] 
\renewcommand{\arraystretch}{2.0}
	\centering
	\caption{\label{table:d1} A table of source and field operators for each term in the stress tensor.}
			\vspace{0.5cm}
		\begin{tabular}{l c c}
		\hline		
		\quad&Field Component  & \quad \quad Source \ \ $(\v r_i = \v r)$ \\ 
		\hline		
		\quad (1) \quad \quad & $E_n(\v r, s_1)$ & \quad  $ \hat{n} {\delta}(\v r'-\v r_i)$  \\ 
		\hline
		\quad (2) \quad & $H_u(\v r, s_2)$ & \quad 	
		 $\hat{v}\ \pd{}{n'} {\delta}(\v r'-\v r_i)$ \\[0ex]	
	\hline 
	   \quad (3)\quad & $H_u(\v r, s_3)$ & \quad 
		 $\hat{n}\ \pd{}{v'} {\delta}(\v r'-\v r_i)$ \\
	\hline 
	   \quad (4)\quad & $H_v(\v r, s_4)$ & \quad 	
		 $\hat{n}\ \pd{}{u'} {\delta}(\v r'-\v r_i)$ \\[0ex]	
	\hline 
	   \quad (5)\quad &$H_v(\v r, s_5)$ & \quad 	
		 $\hat{u}\ \pd{}{n'} {\delta}(\v r'-\v r_i)$ \\
	\hline 
		\end{tabular}
	
\end{table}

%Substituting \eqref{eqn:d7} and \eqref{eqn:d8} into \eqref{eqn:d6}, the normal component of the Casimir force can be expressed as:
%\begin{eqnarray}  \label{eqn:c9}
%%F_n(\v r) = T_{nn}(\v r)  = \frac{\hbar c_0}{2 \pi}\text{Im} \int_0^{\infty} \!\!\! dk \left[ \underbrace{\frac{k^2}{i\omega\mu_0} E_{n}(\v r, s_1)}_{T_1} + \underbrace{H_u(\v r, s_2)}_{T_2} - \underbrace{ H_u(\v r, s_3)}_{T_3}
%%\right. \nonumber \\ \left.+ \underbrace{H_v(\v r, s_4)}_{T_4} -\underbrace{ H_v(\v r, s_5)}_{T_5} \right]_{\v r' = \v r} \hspace{0cm}
%P(\v r) = T_{nn}(\v r)  = \frac{\hbar c_0}{2 \pi}\text{Im} \int_0^{\infty} \!\!\! dk \left[ \frac{k^2}{i\omega\mu_0} E_{n}(\v r, s_1) + {H_u(\v r, s_2)} - { H_u(\v r, s_3)}
%\right. \nonumber \\ \left.+ {H_v(\v r, s_4)} -{ H_v(\v r, s_5)} \right]_{\v r' = \v r} \hspace{0cm}
%\end{eqnarray}
%\subsection{2D Integral Equation for $z$-invariant Structures}

If the object is made of perfect conducting material, there is only induced electric current $\v J$ flowing on the surface. The scattered electric field and magnetic field in the free space region could be obtained from the induced electric current by $\mathcal L$ and $\mathcal K$ operators respectively \cite{PMCHWT}: 
%\vspace{-0.0cm}
\begin{eqnarray}
\v E (\v r) &=& \v E^i(\v r) + \mathcal L \v J(\v r)  \\
\v H (\v r) &=& \v H^i(\v r) - \mathcal K \v J(\v r) 
\end{eqnarray}
where $\v E$ and $\v H$ are the total field in the space and $\v E^i$ and $\v H^i$ are the incident field generated by an external source. The operators are defined as
%\begin{eqnarray}
%\mathcal{L} \v{X}(\v{r}) &=& i\omega\mu_0 \int_s \left[\dyad I + \frac{\nabla\nabla}{k^2_0} \right]\grr \cdot \v{X}(\v{r'})d\v{r'} + \frac{i}{4\pi\omega\epsilon_0}\int_s 
%\nabla \nabla' \cdot [\grr \v{X}(\v{r'})]d\v{r'} \label{eqn:d21} 
%\end{eqnarray}
\begin{eqnarray}
\mathcal{L} \v{X}(\v{r}) &=& i\omega\mu_0 \int_s \left[\dyad I + \frac{\nabla\nabla}{k^2_0} \right]\grr \cdot \v{X}(\v{r'})d\v{r'}  \label{eqn:d21}\\
\mathcal{K} \v{X}(\v{r}) &=& \int_s \v{X}(\v{r'}) \times \nabla \grr d\v{r'}\label{eqn:d22}
\end{eqnarray}
where $g(\prs)  = e^{ik_0|\v r - \v r'|}/4\pi|\v r - \v r'|$. The total tangential electric field vanishes on the surface of the PEC object. Enforcing Eqn. \eqref{eqn:d21} on the surface we obtain the electric field integral equation(EFIE):
\begin{eqnarray}
-\hat n \times \v E^i(\v r) = \hat n \times \mathcal L \v J(\v r) \label{eqn:d23}
\end{eqnarray}
By expanding the unknown current with RWG basis \cite{RWG}, $\v J(\v r') = \sum_n a_n \v J_n(\v r')$ and using Galerkin's method, the integral equation \eqref{eqn:d23} could be cast into a matrix equation $[Z_{mn}][a_n] = [g_m]$, where 
\begin{eqnarray}
g_m &=& \int_{T_m}d \v r \v J_m (\v r) \cdot \v E^i (\v r) \\
Z_{mn} &=& i\omega\mu_0\int_{T_m}d \v r \int_{T_n}d \v r' \v J_m \cdot\left[\dyad I + \frac{\nabla\nabla}{k^2_0} \right] g(\prs)  \cdot \v J_n (\v r')
\end{eqnarray}
If the sources listed in Table \ref{table:d1} are used to generate the incident field, then the evaluation of $g_m$ involves integrals containing some super-hyper singularities. The treatment of these singularities has been discussed in detail in \cite{Meisong_2007}. 

%****************************************************************************
%\section{Numerical Results}

Now we would present some numerical results obtained from this method. They are compared with either analytical results or other's calculation. First we would start with the evaluation of the Casimir force between two spheres, since they are the simplest 3D objects and the analytical expression of the Casimir energy is available in \cite{EGJK_2007}. We assume that the radius of both spheres is $R$ and their minimum distance is $Z$, the product of the force and the square of the radius ($F*R^2/\hbar c_0$) is a dimensionless and scale-invariant quantity. So the result is given by their product ($F*R^2/\hbar c_0$) versus the normalized separation ($Z/R$). In the numerical process, first the Casimir force per unit area (pressure) is calculated at different points on the surface, and then the total force is obtained by integrating the pressure over the entire surface. The analytical results are interpolated from the data given in ref. \cite{EGJK_2007}. Good agreement has been observed. 
%Each sphere is discretized into 286 triangles and each data point can be obtained within 3 to 4 minutes on a 3.0 GHz computer. The maximum relative error is about $5\%$ percent. The average relative error is $3\%$

\begin{figure}[htbp]
\begin{center}
\includegraphics[width=0.45\textwidth]{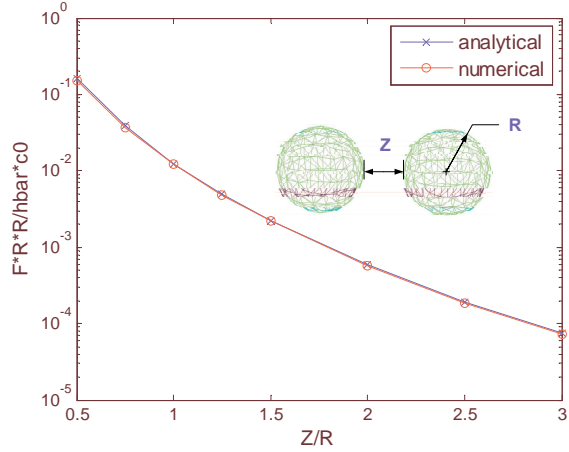}% Here is how to import EPS art
\end{center}
\caption{\label{fig:sphere} Casimir force between two spheres of 
radius $R$ and a separation of $Z$. }
\end{figure}

%Compared to the method presented in \cite{EGJK_2007} and \cite{Reid_2009}, our method has the advantage that we can obtain the pressure distribution on the surface of the object.
% The pressure distribution on one of the spheres in this example is shown in Figure \ref{fig:sphere_dist}. 
%
%\begin{figure}[htbp]
%\begin{center}
%\includegraphics[width=0.7\textwidth]{casimir/sphere_force_dist}% Here is how to import EPS art
%\end{center}
%\caption{\label{fig:sphere_dist} Casimir pressure distribution on the surface of one of the spheres $(Z=R)$. The force is shown in log scale.}
%\end{figure}

The geometry of the second example contains two identical capsules. A capsule is a cylinder of radius $R$ with hemispherical end caps. Its total length is denoted by $L$. Experiment had been carried out to measure the interaction between two crossed cylindrical capsules \cite{Ederth_2000} so a numerical study of the geometry would be interesting to experimentalist. It has also been simulated by using EGJK method in ref. \cite{Reid_2009} and reference data is available to make comparison. The force between two capsules is evaluated for two cases: they are parallel or they are perpendicular. 
%The reference data is obtained by interpolating the numerical results given in ref. \cite{Reid_2009}.
%Each capsule is discretized into 852 triangles. 
\begin{figure}[htbp]
\begin{center}
\includegraphics[width=0.45\textwidth]{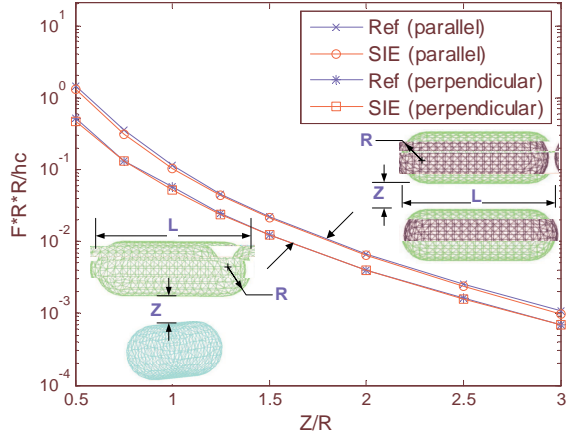}% Here is how to import EPS art
\end{center}
\caption{\label{fig:capsule} Casimir for two pairs 
of perfectly conducting objects: 
identical capsules with parallel axes, and identical
capsules with perpendicular axes. Here $L = 6R$.  SIE refers to the surface integral equation method derived in this paper.}
\end{figure}

%****************************************************************************
%\section{Conclusion}
Similar to the 2D geometries \cite{Jie_2009} case, the integral equation method could reduced the computation cost significantly since it involves only surface unknowns.
Moreover, both the coordinate space integration \eqref{eqn:c1} and the spectral
space integration \eqref{eqn:c33} to \eqref{eqn:c35} are smooth and independent of
the number of unknowns. 

As we know, both our surface integral method following the stress tensor approach and  Reid et al numerical method following the path integral approach \cite{Reid_2009} are effective numerical methods for calculating the Casimir force among arbitrary 3D objects. Though they start from completely different description of the Casimir energy and force, they are connected with the surface integral equation method in classical electromagnetics at certain stage. Reid et al method reduces the problem to a search of the all eigenvalues of the impedance matrix. Our method reduces the problem to solving a set of matrix equation, with the same impedance matrix as in Reid et al method. Reid et al method calculates energy directly while our method calculates the force distribution directly. The complexity of both method could be reduced to $O(N\log N)$ at best by using fast algorithms, where $N$ is the number of surface unknowns. The drawback of the path integral method is that we need to differentiate the total energy with respect to displacement in a certain direction to get the force. This operation changes the problem into a generalized eigenvalue problem which increases the cost and complexity. Another drawback is that the force distribution on the surface is not available from the method. On the contrary, for stress tensor approach, force distribution is required before we obtain the total force. The tradeoff is that an additional surface integral needs to be performed compared to the path integral method. 

%****************************************************************************
% the following vfill coursely balances the columns on the last page
% \vfill
% \pagebreak

%****************************************************************************
%\section*{Acknowledgement}
%The authors wish to acknowledge many helpful discussions with Alejandro W. Rodriguez.
%\newline
%\begin{figure}[htbp]
%	\centering
%		\includegraphics[width=0.45\textwidth]{piston_fig2.eps}
%	\label{fig:piston_fig2}
%\end{figure}

%\bibliography{apssamp}% Produces the bibliography via BibTeX.

\end{document}